\documentclass[12pt]{iopart}

\usepackage{iopams}
\usepackage{braket}
\usepackage{amsfonts}
\usepackage{color}
\usepackage[dvipsnames]{xcolor}
\usepackage{amssymb}
\usepackage{mathrsfs}
\usepackage{graphicx}
\usepackage{lmodern}
\usepackage{lineno}
\usepackage{hyperref}
\usepackage[normalem]{ulem}

\newcommand{\beq}{\begin{eqnarray}}
\newcommand{\eeq}{\end{eqnarray}}

\def\keywords#1{\vspace{10pt}
     \begin{indented}
     \item[]\rm Keywords: #1\par
     \end{indented}}

\begin{document}



\title{Wigner function under changes of reference frames}
\author{J. Berra--Montiel$^{1,2}$ and G. F. Torres del Castillo$^{3}$}

\address{$^{1}$ Facultad de Ciencias, Universidad Aut\'onoma de San Luis 
Potos\'{\i} \\
Campus Pedregal, Av. Parque Chapultepec 1610, Col. Privadas del Pedregal, San
Luis Potos\'{\i}, SLP, 78217, Mexico}
\address{$^2$ Dipartimento di Fisica ``Ettore Pancini", Universit\'a degli studi di Napoli ``Federico II", Complesso Univ. Monte S. Angelo, I-80126 Napoli, Italy}
\address{$^3$Instituto de Ciencias,
Benem\'erita Universidad Aut\'onoma de Puebla, 72570 Puebla, Pue., M\'exico}

\eads{\mailto{\textcolor{blue}{jasel.berra@uaslp.mx}},\
\mailto{\textcolor{blue}{gtorres@fcfm.buap.mx}} 
}


\begin{abstract}

In this paper, we investigate the transformation laws of the Wigner function under changes of reference frames. By employing the coordinate transformation of the wave functions, we derive an integral representation for the transformed Wigner function in both position and momentum representations. To illustrate our results, we include some basic examples.

\end{abstract}

\keywords{Wigner function, Phase Space Quantum Mechanics, Reference frames}


\section{Introduction}
Reference frames are fundamental in physics. Whenever we set up an experiment or describe the behaviour of a physical system, we implicitly or explicitly rely on a reference frame. Broadly speaking, a reference frame is a conceptual or physical framework relative to which we measure the properties of the system under study. The mathematical counterpart of a reference frame is provided by a coordinate system and a time scale, which together allow us to determine the motion of particles. In classical mechanics, a change of reference frame typically involves straightforward transformations of positions and momenta, which do not affect the underlying physical processes, rather, they merely alter the mathematical description. However, in quantum mechanics, the situation becomes more intricate due to the wave-like behaviour of particles and the significant role played by the wave function. For instance, in non-relativistic quantum mechanics the coordinate transformations are usually presented in terms of unitary operators that correspond to canonical transformations in classical phase space \cite{Moshinsky}. Nevertheless, despite considerable efforts over the years, a complete theory of coordinate transformations in quantum mechanics remains elusive, including a satisfactory theory of canonical transformations. \\
\noindent In order to shed some light on these issues, in this paper we analyze the transformation laws under changes of reference frames within the phase space formulation of quantum mechanics. The phase space formulation of quantum mechanics consists in a formal passage from classical to quantum systems using the Dirac quantization framework as a fundamental guideline \cite{DQ}. A central feature within this formulation is determined by the Wigner distribution function. This function provides a phase space representation of the density operator and captures
all auto-correlation properties and transition amplitudes of a quantum system. As we will demonstrate below, the transformation of wave functions enables us to characterize the change of the Wigner function entirely in terms of the transformations of the extended phase space variables.

The paper is organized as follows, in section 2 we briefly introduce the fundamental concepts of phase space quantization and Wigner functions. In section 3, the coordinate transformations of wave functions are analyzed. Then, in section 4 we obtain the transformation laws of the Wigner functions under changes of reference frames, presenting several examples. Finally, we introduce some concluding remarks in section 5.

\section{The Wigner-Weyl quantization and the Wigner function}

The Wigner-Weyl quantization is a formulation of quantum mechanics that represents quantum states and observables as functions on the phase space, offering an alternative to the traditional wavefunction or operator formalism based on Hilbert spaces. In this framework, quantum observables, which typically act as operators on states defined in Hilbert spaces, are mapped to phase space functions through the Weyl transform, establishing a direct correspondence between classical observables and their quantum counterparts \cite{Cosmas}. By avoiding the need for Hilbert space operators, the Wigner-Weyl quantization provides an intuitive approach that proves to be useful for understanding the transition between classical and quantum regimes, as well as for practical applications in areas like quantum optics, quantum information and semiclassical approximations \cite{Overview}.

\noindent Let us consider a classical system with $n$ degrees of freedom described by the phase space $\mathbb{R}^{2n}$, with local coordinates $\mathbf{x}=(x_{1},\ldots, x_{n})$ and $\mathbf{p}=(p_{1},\ldots,p_{n})$. The quantization of this theory results in the construction of the Hilbert space $L^{2}(\mathbb{R}^{n})$, where the functions $\mathbf{x}$ and $\mathbf{p}$ become operators $\hat{\mathbf{x}}=(\hat{x}_{1},\ldots, \hat{x}_{n})$ and $\hat{\mathbf{p}}=(\hat{p}_{1},\ldots,\hat{p}_{n})$, acting on $L^{2}(\mathbb{R}^{n})$, which satisfy the commutation relations
\begin{equation}
[\hat{x}_{i},\hat{x}_{j}]=0, \;\; [\hat{p}_{i},\hat{p}_{j}]=0, \;\; [\hat{x}_{i},\hat{p_{j}}]=i\hbar\delta_{ij}.
\end{equation}
Now, given a function $f(\mathbf{x},\mathbf{p})$ on $\mathbb{R}^{2n}$, we can promote this classical function to an operator through the Weyl quantization map, defined as
\begin{equation}\label{Weyl}
\mathcal{Q}(f)=\frac{1}{(2\pi)^{n}}\int_{\mathbb{R}^{2n}}\tilde{f}(\mathbf{a},\mathbf{b})e^{i(\mathbf{a}\cdot\hat{\mathbf{x}}+\mathbf{b}\cdot{\hat{\mathbf{p}}})}\,d\mathbf{a}\,d\mathbf{b},
\end{equation} 
where $\tilde{f}$ denotes the Fourier transform of $f$. Since the Weyl transform corresponds to the integral of an operator, we can compute its integral kernel as 
\begin{equation}
K_{f}(\mathbf{x},\mathbf{x}')=\braket{\mathbf{x}|\mathcal{Q}(f)|\mathbf{x}'}.
\end{equation} 
By means of the Baker-Campbell-Hausdorff formula, we obtain
\begin{equation}
K_{f}(\mathbf{x},\mathbf{x}')=\frac{1}{(2\pi\hbar)^{n}}\int_{\mathbb{R}^{n}}f\left(\frac{\mathbf{x}+\mathbf{x}'}{2},\mathbf{p} \right)e^{\frac{i}{\hbar}\mathbf{p}\cdot(\mathbf{x}-\mathbf{x}')}\,d\mathbf{p}. 
\end{equation}
This integral kernel for the operator $\mathcal{Q}(f)$ also satisfies
\begin{equation}
\int_{\mathbb{R}^{2n}}\left| K_{f}(\mathbf{x},\mathbf{x}')\right|^{2}\,d\mathbf{x}\,d\mathbf{x}'=\frac{1}{(2\pi\hbar)^{n}}\int_{\mathbb{R}^{2n}}\left| f(\mathbf{x},\mathbf{p})\right|^{2}\,d\mathbf{x}\,d\mathbf{p},
\end{equation}
which implies that the kernel defines a Hilbert-Schmidt operator, as it maps functions from $L^{2}(\mathbb{R}^{n})$ into functions in $L^{2}(\mathbb{R}^{n})$ and possesses a well defined (but possibly infinite) trace \cite{Reed}. Then, we conclude that the Weyl transform (\ref{Weyl}) can be viewed as a map from the set of square-integrable functions on the phase space $\mathbb{R}^{2n}$ to the space of Hilbert-Schmidt operators $\mathrm{HS}(L^{2}(\mathbb{R}^{n}))$ acting on $L^{2}(\mathbb{R}^{n})$. In particular, it can be shown that the Weyl transform $\mathcal{Q}(f)$ corresponds to a self-adjoint operator when $f$ is a real-valued function \cite{Takhtajan}.

\noindent Once we have described the quantization map, the inverse of the Weyl transform $\mathcal{Q}^{-1}:\mathrm{HS}(L^{2}(\mathbb{R}^{n}))\to L^{2}(\mathbb{R}^{2n})$ can be obtained as follows
\begin{equation}\label{WignerT}
\mathcal{Q}^{-1}(\hat{A})=A_{W}(\mathbf{x},\mathbf{p})=\int_{\mathbb{R}^{n}}\braket{\mathbf{x}-\frac{\mathbf{x}'}{2}|\hat{A}|\mathbf{x}+\frac{\mathbf{x}'}{2}}e^{\frac{i}{\hbar}\mathbf{p}\cdot\mathbf{x}'}\,d\mathbf{x}',
\end{equation}     
for any operator $\hat{A}\in \mathrm{HS}(L^{2}(\mathbb{R}^{n}))$. The resulting function $A_{W}(\mathbf{x},\mathbf{p})$, defined on the phase space $\mathbb{R}^{2n}$, is known as the Wigner transform of the operator $\hat{A}$. Although these expressions were derived for square-integrable functions in phase space, they can be extended to the case of arbitrary functions through the use of generalized functions \cite{Reed}.

\noindent With the Weyl quantization map and the Wigner transform established, we are now ready to define the Wigner function or Wigner distribution associated to the wavefunction $\psi(\mathbf{q})$ belonging to the Hilbert space $L^{2}(\mathbb{R}^{n})$. Consider $\hat{\rho}$, a density operator corresponding to a quantum state $\psi\in L^{2}(\mathbb{R}^{n})$, that is, a self-adjoint, positive and semi-definite operator with trace one, written as
\begin{equation}
\hat{\rho}_{\psi}=\ket{\psi}\bra{\psi},
\end{equation}    
in Dirac notation. By means of the Wigner transform formula (\ref{WignerT}), the phase space function associated to the state $\psi$, also known as the Wigner function, is given by
\begin{equation}
\hspace{-4em} W(\mathbf{x},\mathbf{p})=\frac{1}{(2\pi\hbar)^{n}}\mathcal{Q}^{-1}(\hat{\rho})=\frac{1}{(2\pi\hbar)^{n}}\int_{\mathbb{R}^{n}}\psi^{*}\left(\mathbf{x}+\frac{\mathbf{x}'}{2}\right) \psi\left(\mathbf{x}-\frac{\mathbf{x}'}{2}\right)e^{\frac{i}{\hbar}\mathbf{p}\cdot\mathbf{x}'}\,d\mathbf{x}', 
\end{equation}
where the extra overall factor $1/(2\pi\hbar)^{n}$ has been chosen so that the Wigner function turns out to be normalized in phase space 
\begin{equation}
\int_{\mathbb{R}^{2n}}W(\mathbf{x},\mathbf{p})\,d\mathbf{x}\,d\mathbf{p}=1.
\end{equation} 
A remarkable feature of the Wigner function for a quantum state lies on the possibility of taking negative values on certain regions of phase space. Consequently, it cannot be interpreted as a probability density in the traditional sense, and is therefore commonly referred to as a quasi-probability distribution in the literature. However, this seemingly unusual property of the Wigner function enables the characterization of joint-correlation functions and entanglement properties within the quantum system \cite{Recent}. Moreover, the Wigner function contains all the necessary information about the quantum probability distributions associated with a state. Indeed, it is straightforward to verify from its definition that
\begin{equation}
\int_{\mathbb{R}^{2n}}W(\mathbf{x},\mathbf{p})\,d\mathbf{p}=|\psi(\mathbf{x})|^{2},
\end{equation}
which corresponds to the probability distribution in position space, and similarly, for the momentum space one finds
\begin{equation}
\int_{\mathbb{R}^{2n}}W(\mathbf{x},\mathbf{p})\,d\mathbf{x}=|\tilde{\psi}(\mathbf{p})|^{2},
\end{equation}
where $\tilde{\psi}(\mathbf{p})$ denotes the Fourier transform of the wavefunction $\psi(\mathbf{x})$. 
The Wigner function can also be employed to calculate the expectation values of operators by integrating their respective Wigner transforms across the entire phase space,
\begin{equation}
\braket{\psi|\hat{A}|\psi}=\int_{\mathbb{R}^{2n}}W(\mathbf{x},\mathbf{p})A_{W}(\mathbf{x},\mathbf{p})\,d\mathbf{x}\,d\mathbf{p}.
\end{equation}
All these properties suggest that the Wigner function is the closest analogue to a probability distribution in a quantum system. Since it represents the phase space counterpart of the density operator, one could argue that the information contained in the Wigner function is fully equivalent to that provided by the wavefunctions in the conventional formulation of quantum mechanics \cite{Curtright}.

\section{Transformation of wavefunctions}
Following \cite{Torres}, let us consider a transformation given by a unitary operator $\hat{U}$, defined by the conditions
\begin{equation}\label{Unitary}
\hat{U}\hat{x_{i}}\hat{U}^{-1}=\hat{X}_{i}(\hat{\mathbf{x}},t), \;\;\;\hat{U}\hat{p_{i}}\hat{U}^{-1}=\hat{P}_{i}(\hat{\mathbf{p}},t),
\end{equation}
where $\hat{x}_{i}$ and $\hat{p}_{j}$ denote the position and momentum operators respectively, while $\hat{X}_{i}$ and $\hat{P}_{j}$ are operators depending on $\hat{\mathbf{x}}$, $\hat{\mathbf{p}}$ and $t$. Furthermore, within the Schr\"odinger picture, the states transform under the action of unitary operators as follows
\begin{equation}\label{Transformstate}
\ket{\psi'}=\hat{U}\ket{\psi}.
\end{equation} 
Now, suppose that $\ket{\mathbf{x}}$ and $\ket{\mathbf{p}}$ are eigenstates of the position and momentum operators $\hat{\mathbf{x}}$ and $\hat{\mathbf{p}}$, respectively, such that $\hat{\mathbf{x}}\ket{\mathbf{x}}=\mathbf{x}\ket{\mathbf{x}}$ and $\hat{\mathbf{p}}\ket{\mathbf{p}}=\mathbf{p}\ket{\mathbf{p}}$, then by using Eqs. (\ref{Unitary}) the following result holds
\begin{equation}
\hat{\mathbf{x}}\hat{U}^{-1}\ket{\mathbf{x}}=\hat{U}^{-1}\hat{\mathbf{X}}(\hat{\mathbf{x}},t)\ket{\mathbf{x}}=\mathbf{X}(\mathbf{x},t)\hat{U}^{-1}\ket{\mathbf{x}},
\end{equation} 
which implies that $\hat{U}^{-1}\ket{\mathbf{x}}$ is a eigenstate of the operator $\hat{\mathbf{x}}$ with eigenvalue $\mathbf{X}(\mathbf{x},t)$, thus
\begin{equation}\label{Ux}
\hat{U}^{-1}\ket{\mathbf{x}}=e^{\frac{i}{\hbar}\alpha(\mathbf{x},t)}\ket{\mathbf{X}(\mathbf{x},t)},
\end{equation}
where, in general, $\alpha(\mathbf{x},t)$ corresponds to a real parameter depending on $\mathbf{x}$ and $t$. According to  equation (\ref{Transformstate}), the former expression allows us to write the transformed wave functions as follows
\begin{eqnarray}\label{Transwave}
\psi'(\mathbf{x})&=&\braket{\mathbf{x}|\hat{U}|\psi}=e^{-\frac{i}{\hbar}\alpha(\mathbf{x},t)}\braket{\mathbf{X}(\mathbf{x},t)|\psi} \nonumber \\
&=& e^{-\frac{i}{\hbar}\alpha(\mathbf{x},t)}\psi\left( \mathbf{X}(\mathbf{x},t)\right). 
\end{eqnarray} 
A similar argument shows that 
\begin{equation}\label{Up}
\hat{U}^{-1}\ket{\mathbf{p}}=e^{\frac{i}{\hbar}\beta(\mathbf{p},t)}\ket{\mathbf{P}(\mathbf{p},t)},
\end{equation}
where $\beta(\mathbf{p},t)$ is a real parameter that depends on $\mathbf{p}$ and $t$. In order to determine the values of the parameters $\alpha$ and $\beta$, we calculate the scalar product
\begin{equation}
\braket{\mathbf{x}|\hat{U}\hat{U}^{-1}|\mathbf{p}}=\braket{\mathbf{x}|\mathbf{p}}=\frac{1}{(2\pi\hbar)^{n/2}}e^{\frac{i}{\hbar}\mathbf{p}\cdot\mathbf{x}},
\end{equation}
then, by applying the unitary transformations (\ref{Ux}) and (\ref{Up}), the former equation must agree with
\begin{equation}
\hspace{-1em}e^{\frac{i}{\hbar}\left( \beta(\mathbf{p},t)-\alpha(\mathbf{x},t)\right)}\braket{\mathbf{X(\mathbf{x},t)}|\mathbf{P}(\mathbf{p},t)}=\frac{1}{(2\pi\hbar)^{n/2}}e^{\frac{i}{\hbar}\left( \beta(\mathbf{p},t)-\alpha(\mathbf{x},t)+\mathbf{P}(\mathbf{p},t)\cdot\mathbf{X}(\mathbf{x},t)\right)}.
\end{equation}
This implies that
\begin{equation}\label{alpha}
\mathbf{p}\cdot{\mathbf{x}}=\beta(\mathbf{p},t)-\alpha(\mathbf{x},t)+\mathbf{P}(\mathbf{p},t)\cdot\mathbf{X}(\mathbf{x},t).
\end{equation}
As we will see in the next section, the preceding formula enable us to describe the transformation of the Wigner function under unitary transformations associated with changes of reference frames.  

\section{Transformation of the Wigner function under changes of reference frames}
In order to determine the behaviour of the Wigner function under the action of unitary transformations defined by the conditions (\ref{Unitary}), let us write the Wigner function associated to the density operator $\hat{\rho}_{\psi'}=\ket{\psi'}\bra{\psi'}$ for the state $\psi' \in L^{2}(\mathbb{R}^{n})$ as
\begin{equation}
W'(\mathbf{x},\mathbf{p})=\frac{1}{(2\pi\hbar)^{n}}\int_{\mathbb{R}^{n}}\braket{\mathbf{x}-\frac{\mathbf{y}}{2}|\psi'}\braket{\psi'|\mathbf{x}+\frac{\mathbf{y}}{2}}e^{\frac{i}{\hbar}\mathbf{p}\cdot\mathbf{y}}\,d\mathbf{y}.
\end{equation}  
Now, by making use of the action of the unitary operators on the states (\ref{Transformstate}) and the transformation of the wavefunctions (\ref{Transwave}), we have
\begin{eqnarray}\label{TWignerq}
 W'(\mathbf{x},\mathbf{p})&=&\frac{1}{(2\pi\hbar)^{n}}\int_{\mathbb{R}^{n}}\braket{\mathbf{x}-\frac{\mathbf{y}}{2}|\hat{U}|\psi}\braket{\psi|\hat{U}^{-1}|\mathbf{x}+\frac{\mathbf{y}}{2}}e^{\frac{i}{\hbar}\mathbf{p}\cdot\mathbf{y}}\,d\mathbf{y}, \nonumber \\
&=& \frac{1}{(2\pi\hbar)^{n}}\int_{\mathbb{R}^{n}}\psi^{*}\left(\mathbf{X}\left(\mathbf{x}+\frac{\mathbf{y}}{2},t \right)\right)\psi\left(\mathbf{X}\left(\mathbf{x}-\frac{\mathbf{y}}{2},t\right)\right) \nonumber \\
&& \times e^{-\frac{i}{\hbar}\left[ \alpha\left( \mathbf{x}-\frac{\mathbf{y}}{2},t\right) -\alpha\left( \mathbf{x}+\frac{\mathbf{y}}{2},t\right)-\mathbf{p}\cdot\mathbf{y}\right]}\,d\mathbf{y}.
\end{eqnarray}
Similarly, through a Fourier transform, an analogous expression for the Wigner function in the momentum representation holds
\begin{equation}
W'(\mathbf{x},\mathbf{p})=\frac{1}{(2\pi\hbar)^{n}}\int_{\mathbb{R}^{n}}\braket{\mathbf{p}-\frac{\mathbf{u}}{2}|\psi'}\braket{\psi'|\mathbf{p}+\frac{\mathbf{u}}{2}}e^{-\frac{i}{\hbar}\mathbf{x}\cdot\mathbf{u}}\,d\mathbf{u}.
\end{equation}  
Using in this case equation (\ref{Up}), we can derive    
 \begin{eqnarray}\label{TWignerp}
W'(\mathbf{x},\mathbf{p})&=&\frac{1}{(2\pi\hbar)^{n}}\int_{\mathbb{R}^{n}}\braket{\mathbf{p}-\frac{\mathbf{u}}{2}|\hat{U}|\psi}\braket{\psi|\hat{U}^{-1}|\mathbf{p}+\frac{\mathbf{u}}{2}}e^{-\frac{i}{\hbar}\mathbf{x}\cdot\mathbf{u}}\,d\mathbf{u}, \nonumber \\
&=& \frac{1}{(2\pi\hbar)^{n}}\int_{\mathbb{R}^{n}}\tilde{\psi}^{*}\left(\mathbf{P}\left(\mathbf{p}+\frac{\mathbf{u}}{2},t \right)\right)\tilde{\psi}\left(\mathbf{P}\left(\mathbf{p}-\frac{\mathbf{u}}{2},t\right)\right) \nonumber \\
&& \times e^{\frac{i}{\hbar}\left[ \beta\left( \mathbf{p}+\frac{\mathbf{u}}{2},t\right) -\beta\left( \mathbf{p}-	\frac{\mathbf{u}}{2},t\right)-\mathbf{x}\cdot\mathbf{u}\right]}\,d\mathbf{u}.
\end{eqnarray}
As we will observe in the following subsections, the inclusion of the parameters $\alpha$ and $\beta$ proves to be essential in order to express the transformed Wigner function entirely in terms of the new position and momentum coordinates. 

\subsection{Examples}
In this subsection, we present some well-known examples with the main aim to explicitly illustrate how the Wigner function is transformed under a change of reference frame using the method outlined above.

\subsubsection{Spatial translations.}
As a first test case, let us consider a spatial translation by a constant vector $\mathbf{a}\in\mathbb{R}^{n}$, this transformation can be defined by
\begin{equation}
\hat{\mathbf{X}}=\hat{\mathbf{x}}-\mathbf{a}, \;\;\; \hat{\mathbf{P}}=\hat{\mathbf{p}}.
\end{equation} 
From equation (\ref{alpha}) we obtain $\mathbf{p}\cdot\mathbf{x}=\beta(\mathbf{p},t)-\alpha(\mathbf{x},t)+\mathbf{p}\cdot(\mathbf{x}-\mathbf{a})$, which implies that
\begin{equation}
\alpha(\mathbf{x},t)=\beta(\mathbf{p},t)-\mathbf{p}\cdot\mathbf{a}.
\end{equation}
Hence, we conclude that
\begin{equation}
\alpha(\mathbf{x},t)=\xi(t), \;\;\; \beta(\mathbf{p},t)=\mathbf{p}\cdot\mathbf{a}+\xi(t),
\end{equation}
where $\xi(t)$ denotes a real-valued function depending on $t$. When the Hamiltonian is invariant under the transformation (\ref{Unitary}), the function $\xi$ can be chosen in such a way that the transformation (\ref{Transwave}) maps solutions of the corresponding Schr\"odinger equation into solutions of this equation, see Ref. \cite{Torres}. Now, substituting the expression for $\alpha$ into equation (\ref{TWignerq}), we obtain
\begin{eqnarray}
 W'(\mathbf{x},\mathbf{p})&=&\frac{1}{(2\pi\hbar)^{n}}\int_{\mathbb{R}^{n}}\psi^{*}\left(\mathbf{x}-\mathbf{a}+\frac{\mathbf{y}}{2}\right) \psi\left(\mathbf{x}-\mathbf{a}-\frac{\mathbf{y}}{2}\right)e^{\frac{i}{\hbar}\mathbf{p}\cdot\mathbf{y}}\,d\mathbf{y}, \nonumber \\
 &=&\frac{1}{(2\pi\hbar)^{n}}\int_{\mathbb{R}^{n}}\psi^{*}\left(\mathbf{X}+\frac{\mathbf{y}}{2}\right) \psi\left(\mathbf{X}-\frac{\mathbf{y}}{2}\right)e^{\frac{i}{\hbar}\mathbf{P}\cdot\mathbf{y}}\,d\mathbf{y}, \nonumber \\
 &=& W(\mathbf{X},\mathbf{P}).
\end{eqnarray}
Similarly, we can make use of the parameter $\beta$ and the momentum representation (\ref{TWignerp}) to obtain the transformed Wigner function as
\begin{eqnarray}
 W'(\mathbf{x},\mathbf{p})&=&\frac{1}{(2\pi\hbar)^{n}}\int_{\mathbb{R}^{n}}\tilde{\psi}^{*}\left(\mathbf{p}+\frac{\mathbf{u}}{2}\right) \tilde{\psi}\left(\mathbf{p}-\frac{\mathbf{u}}{2}\right))e^{-\frac{i}{\hbar}(\mathbf{x}-\mathbf{a})\cdot\mathbf{u}}\,d\mathbf{u}, \nonumber \\
 &=& \frac{1}{(2\pi\hbar)^{n}}\int_{\mathbb{R}^{n}}\tilde{\psi}^{*}\left(\mathbf{P}+\frac{\mathbf{u}}{2}\right)\tilde{\psi}\left(\mathbf{P}-\frac{\mathbf{u}}{2}\right)e^{-\frac{i}{\hbar}\mathbf{X}\cdot\mathbf{u}}\,d\mathbf{u}, \nonumber \\
 &=& W(\mathbf{X},\mathbf{P}).
\end{eqnarray}

\subsubsection{Galilean transformations.} Our next example is devoted to the case of the Galilean transformations, given by
\begin{equation}
\hat{\mathbf{X}}=\hat{\mathbf{x}}-\mathbf{V}t, \;\;\; \hat{\mathbf{P}}=\hat{\mathbf{p}}-m\mathbf{V},
\end{equation} 
where $\mathbf{V}$ denotes a constant vector in $\mathbb{R}^{n}$. Following \cite{Torres}, in this case, equation (\ref{alpha}) becomes
\begin{equation}
\mathbf{p}\cdot\mathbf{x}=\beta(\mathbf{p},t)-\alpha(\mathbf{x},t)+(\mathbf{p}-m\mathbf{V})\cdot(\mathbf{x}-\mathbf{V}t),
\end{equation}
so that
\begin{eqnarray}
\alpha(\mathbf{x},t)&=&-m\mathbf{V}\cdot\mathbf{x}+\xi(t), \nonumber \\
\beta(\mathbf{p},t)&=&\mathbf{p}\cdot\mathbf{V}t+\xi(t),
\end{eqnarray}
where $\xi(t)$ denotes a real-valued function depending on $t$. From the formula for the transformation of the Wigner function (\ref{TWignerq}), we obtain 
\begin{eqnarray}
 W'(\mathbf{x},\mathbf{p})&=&\frac{1}{(2\pi\hbar)^{3}}\int_{\mathbb{R}^{3}}\psi^{*}\left(\mathbf{x}-\mathbf{V}t+\frac{\mathbf{y}}{2}\right) \psi\left(\mathbf{x}-\mathbf{V}t-\frac{\mathbf{y}}{2}\right)e^{\frac{i}{\hbar}(\mathbf{p}-m\mathbf{V})\cdot\mathbf{y}}\,d\mathbf{y}, \nonumber \\
 &=&\frac{1}{(2\pi\hbar)^{3}}\int_{\mathbb{R}^{3}}\psi^{*}\left(\mathbf{X}+\frac{\mathbf{y}}{2}\right) \psi\left(\mathbf{X}-\frac{\mathbf{y}}{2}\right)e^{\frac{i}{\hbar}\mathbf{P}\cdot\mathbf{y}}\,d\mathbf{y}, \nonumber \\
 &=& W(\mathbf{X},\mathbf{P}).
\end{eqnarray}
In the same manner, an equivalent result can be derived from the momentum representation by using the parameter $\beta$ in place of $\alpha$. 

\subsubsection{Constant acceleration.} Let us consider now the transformation given by a constant acceleration, which corresponds to 
\begin{equation}
\hat{\mathbf{X}}=\hat{\mathbf{x}}-\frac{1}{2}\mathbf{a}t^{2}, \;\;\; \hat{\mathbf{P}}=\hat{\mathbf{p}}-m\mathbf{a}t,
\end{equation} 
where $\mathbf{a}$ is a constant vector in $\mathbb{R}^{3}$. For this example, expression (\ref{alpha}) shows to be
\begin{equation}
\mathbf{p}\cdot\mathbf{x}=\beta(\mathbf{p},t)-\alpha(\mathbf{x},t)+(\mathbf{p}-m\mathbf{a}t)\cdot(\mathbf{x}-\frac{1}{2}\mathbf{a}t^{2}),
\end{equation}
which leads to
 \begin{eqnarray}
\alpha(\mathbf{x},t)&=&-m\mathbf{a}t\cdot\mathbf{x}+\frac{1}{6}ma^{2}t^{3}+\xi(t), \nonumber \\
\beta(\mathbf{p},t)&=&\frac{1}{2}\mathbf{p}\cdot\mathbf{a}t^{2}-\frac{1}{3}ma^{2}t^{3}+\xi(t),
\end{eqnarray}
where $\xi(t)$ denotes a real-valued function on $t$. Then, with the parameter $\alpha$ at hand, the transformation of the Wigner function reads
  \begin{eqnarray}
 W'(\mathbf{x},\mathbf{p},t)&=&\frac{1}{(2\pi\hbar)^{n}}\int_{\mathbb{R}^{n}}\psi^{*}\left(\mathbf{x}-\frac{1}{2}\mathbf{a}t^{2}+\frac{\mathbf{y}}{2}\right) \psi\left(\mathbf{x}-\frac{1}{2}\mathbf{a}t^{2}-\frac{\mathbf{y}}{2}\right)e^{\frac{i}{\hbar}(\mathbf{p}-m\mathbf{a}t)\cdot\mathbf{y}}\,d\mathbf{y}, \nonumber \\
 &=&\frac{1}{(2\pi\hbar)^{n}}\int_{\mathbb{R}^{n}}\psi^{*}\left(\mathbf{X}+\frac{\mathbf{y}}{2}\right) \psi\left(\mathbf{X}-\frac{\mathbf{y}}{2}\right)e^{\frac{i}{\hbar}\mathbf{P}\cdot\mathbf{y}}\,d\mathbf{y}, \nonumber \\
 &=& W(\mathbf{X},\mathbf{P}).
\end{eqnarray}

\noindent It is important to emphasized that the transformation laws of the Wigner functions, as illustrated in the previous examples, take precisely this form since the unitary operators associated to these transformations \cite{Torres}, are given by displacement operators in both position and momentum, accompanied by a phase factor \cite{Knight}. In the general case, it is expected that the transformed Wigner function takes a different form due to the presence of non-linear terms of the position and momentum operators. This complexity arises, for example in the case of non-inertial frames and quantum reference frames, where the unitary operators introduce dependencies that deviate from the linear behavior characteristic of inertial frames.
 
\section{Conclusions}
\label{sec:conclu}

In this paper we have analyzed the transformation laws of the Wigner function under changes of reference frames within the phase space formulation of quantum mechanics. In particular, we have included some physically motivated examples in order to illustrate our approach. It is worthwhile to note that the method developed here can be applied also to more general unitary operators, such as nonlinear transformations and non-inertial changes of reference frames. As a future work, we pretend to implement our approach to the recently introduced concept of quantum reference frames, which generalizes coordinate systems intended to describe states, measurements and dynamical evolution by means of quantum physical systems. This approach suggest that properties such as entanglement and superposition become frame-dependent features \cite{QRF}. Our intention is to make use of the phase space representation of quantum mechanics to investigate these features in more general scenarios. This investigation will be addressed in future work.

\section*{Acknowledgments}
The authors would like to acknowledge support from SNII CONAHCYT-Mexico. JBM thanks to the Dipartimento di Fisica "Ettore Pancini" for the kind invitation and its generous hospitality.

\section*{References}

\bibliographystyle{unsrt}

\end{document}